\documentclass[aps,prl,twocolumn,showpacs]{revtex4}

\usepackage{dcolumn}
\usepackage{bm}

\newcommand{\ls}{\ensuremath{l_s}} 

\def\p{\partial}

\def\slash#1{\ensuremath{\;/\!\!\!\! #1}}

\newcommand{\cF}{{\mathcal{F}}}

\newcommand{\cL}{\mathcal{L}}

\newcommand{\cN}{{\mathcal{N}}}
\newcommand{\cO}{{\mathcal{O}}}

\newcommand{\bS}{{\mathbf{S}}}

\newcommand{\tret}{{t_{\mbox{\scriptsize ret}}}}

\newcommand{\tx}{{\tilde{x}}}
\newcommand{\ttau}{{\tilde{\tau}}}
\newcommand{\tPi}{{\tilde{\Pi}}}

\begin{document}

\title{Generalized Lorentz-Dirac Equation for a Strongly-Coupled
Gauge Theory}

\author{Mariano Chernicoff}
\author{J.~Antonio Garc\'{\i}a}
\author{Alberto
G\"uijosa}
\affiliation{Departamento de F\'{\i}sica de Altas Energ\'{\i}as,
Instituto de Ciencias Nucleares, Universidad Nacional
Aut\'onoma de M\'exico, Apdo.~Postal 70-543, M\'exico D.F. 04510}



\begin{abstract}
We derive a semiclassical equation of motion for a `composite' quark in strongly-coupled large-$N_c$
$\cN=4$ super-Yang-Mills, making use of the AdS/CFT correspondence. The resulting non-linear equation
 incorporates radiation damping, and reduces to the standard Lorentz-Dirac equation for external forces that are small on the scale of the quark Compton wavelength, but has no self-accelerating or pre-accelerating solutions. {}From this equation one can read off a non-standard
dispersion relation for the quark, as well as a Lorentz covariant formula for its radiation rate.
\end{abstract}

\maketitle

Energy conservation implies that a radiating charge must experience a damping
force, originating from its self-field. In the context of
classical electrodynamics, this effect is incorporated in the
(Abraham-)Lorentz-Dirac equation \cite{dirac},
\begin{equation}\label{ald}
m\left({d^2 x^{\mu}\over d\tau^2}-r_e\left[
{d^3 x^{\mu}\over d\tau^3}-{d^2 x_{\nu}\over d\tau^2}\,
{d^2 x^{\nu}\over d\tau^2}{dx^{\mu}\over d\tau}\right]\right)=\cF^{\mu}~,
\end{equation}
with
$\tau$ the proper time and $\cF^{\mu}\equiv\gamma(\vec{F}\cdot\vec{v},\vec{F})$
the four-force.  The electron here is modeled as a
vanishingly small spherically symmetric charge distribution.
The characteristic time/size associated with radiation damping, $r_e\equiv 2e^2/3m$, is set by the classical electron radius.
The second term within the square brackets
is the negative of the rate at which four-momentum is carried away from the charge by radiation, so it is only this term that can properly be called radiation reaction. The first term within the square brackets, usually called the Schott term, is known to arise from the effect of the charge's `bound' (as opposed to radiation) field \cite{teitelboim,rohrlich}.

The appearance of a third-order term in (\ref{ald}) leads to unphysical
behavior, including pre-accelerating and self-accelerating (or `runaway')  solutions. These deficiencies are known to
originate from the assumption that the charge is pointlike. For a charge distribution of
small but finite
size $l$, the above equation is corrected by an infinite number of higher-derivative terms $(l
d/dt)^n$, and is physically sound as long as $l>r_e$ \cite{rohrlich,jackson}.
Upon shifting attention to the quantum case, one intuitively expects the pointlike `bare' electron to acquire an effective size of order the Compton wavelength $\lambda_C\equiv 1/m$, due to its surrounding cloud of virtual particles. Indeed, in \cite{monizsharp} it was shown non-relativistic QED leads to
a generalization of (the non-relativistic version of) (\ref{ald}) where the charge develops a characteristic size $l=\lambda_C$.

Going further to the quantum non-Abelian case is a serious
challenge.
Nevertheless, it is the purpose of this letter to show that the AdS/CFT correspondence
\cite{malda,magoo}
 allows us to address this question rather easily in certain
strongly-coupled non-Abelian gauge theories. In the context of this duality, the quark corresponds to the endpoint of a string, whose body codifies the profile of the non-Abelian (bound and radiation) fields sourced by the quark, including, as we will see, the effect of radiation damping.
 We expect this basic story to apply generally to all instances of the gauge/string duality (including cases with finite temperature or chemical potentials), but
for simplicity we will concentrate on the best understood example:
quark motion in the vacuum of $\cN=4$ super-Yang-Mills (SYM). Besides the gauge field, this maximally supersymmetric and conformally invariant theory contains 6 real scalar fields and 4 Weyl fermions, all in the \emph{adjoint} representation of the gauge group.

It is by now well-known that  $\cN=4$ $SU(N_c)$ SYM with
coupling $g_{YM}$ is, despite appearances, completely equivalent \cite{malda} to Type IIB string theory
on a background that asymptotically approaches the five-dimensional \cite{foot} anti-de Sitter (AdS) geometry
\begin{equation}\label{metric}
ds^2=G_{MN}dx^M dx^N={R^2\over z^2}\left(
-dt^2+d\vec{x}^2+{dz^2}\right)~,
\end{equation}
where ${R^4/\ls^4}=g_{YM}^2 N_c\equiv\lambda$ denotes the 't Hooft coupling, and $\ls$ is the string length. The
radial direction $z$ is mapped holographically into a variable length scale in the gauge
theory, in such a way that $z\to 0$ and $z\to\infty$ are respectively the ultraviolet and infrared limits \cite{uvir}. The directions $x^{\mu}\equiv(t,\vec{x})$ are parallel to the AdS boundary $z=0$ and are
directly identified with the gauge theory directions.
The state of IIB string theory described by the unperturbed metric (\ref{metric}) corresponds to the vacuum of the $\cN=4$ SYM theory, and the closed string sector describing (small or large) fluctuations on top of it fully captures the gluonic ($+$ adjoint scalar and fermionic) physics. The string theory description is under calculational control only for small string coupling and low curvatures, which translates into $N_c\gg 1$,  $\lambda\gg 1$.

{}It is also known that one can add to SYM $N_f$ flavors of matter in the \emph{fundamental}
representation of the $SU(N_c)$ gauge group
by introducing an open string sector associated with a stack of $N_f$ D7-branes \cite{kk}.
 We will refer to these degrees of freedom
as `quarks,' even though, being $\cN=2$ supersymmetric,
they include
both spin $1/2$ and spin $0$ fields.  For $N_f\ll N_c$, the backreaction of the D7-branes on the geometry can be neglected; in the field theory this corresponds to
a `quenched' approximation.

An isolated quark of mass $m$ is dual to
an open string that extends radially from the location
\begin{equation}\label{zm}
z_m={\sqrt{\lambda}\over 2\pi m}~
\end{equation}
on the D7-branes to the AdS horizon at $z\to\infty$. The string dynamics is governed by the Nambu-Goto action
\begin{equation}\label{nambugoto}
S_{\mbox{\scriptsize NG}}=-{1\over 2\pi\ls^2}\int
d^2\sigma\,\sqrt{-\det{g_{ab}}}\equiv \int
d^2\sigma\,\cL_{\mbox{\scriptsize NG}}~, 
\end{equation}
where $g_{ab}\equiv\p_a X^M\p_b X^N G_{MN}(X)$ ($a,b=0,1$) denotes
the induced metric on the worldsheet.
We can exert an external force $\vec{F}$ on the string endpoint by turning on an electric field $F_{0i}=F_i$ on the D7-branes. This amounts to adding to (\ref{nambugoto}) the
usual minimal coupling, which in terms of the endpoint/quark worldline
$x^{\mu}(\tau)\equiv X^{\mu}(\tau,z_m)$ reads
\begin{equation}\label{externalforce}
S_{\mbox{\scriptsize F}}=\int
d\tau\,A_{\mu}(x(\tau))\,{dx^{\mu}(\tau)\over d\tau}~.
\end{equation}

Notice that the string is being described (as is customary) in first-quantized
language, and, as long as it is sufficiently heavy, we are allowed to treat it semiclassically.
In gauge theory language, then,
we are coupling a first-quantized quark to the gluonic ($+$ other SYM) field(s), and then carrying out the full path integral over the strongly-coupled field(s) (the result of which is codified by the AdS spacetime), but treating the path integral over the quark trajectory $x^{\mu}(\tau)$ in a saddle-point approximation.

Variation of  $S_{\mbox{\scriptsize NG}}+S_{\mbox{\scriptsize F}}$ implies the standard Nambu-Goto equation of motion for all interior points of the string, plus the boundary condition
\begin{equation}\label{stringbc}
\Pi^{z}_{\mu}(\tau)|_{z=z_m}=\cF_{\mu}(\tau)\quad\forall~\tau~,
\end{equation}
where $\Pi^{z}_{\mu}\equiv {\p\cL_{\mbox{\scriptsize NG}}}/{\p(\p_z X^{\mu})}$
is the worldsheet (Noether) current associated with spacetime momentum, and
$\cF_{\mu}=-F_{\nu\mu}\p_{\tau}x^{\nu}
=(-\gamma\vec{F}\cdot\vec{v},\gamma\vec{F})$
the Lorentz four-force.

For the interpretation of our results it will be crucial to keep in mind that the quark described by this string is not `bare' but `composite' or `dressed'. This can be seen most clearly by working out the expectation value of the gluonic field surrounding a static quark located at the origin.
For $m\to\infty$ ($z_m\to 0$), the result is just the Coulombic field expected (by conformal invariance) for a pointlike charge \cite{dkk}. For finite $m$ the profile is still Coulombic far away from the origin but in fact becomes non-singular at the location of the quark \cite{martinfsq}. The characteristic thickness of the implied non-Abelian charge distribution
 is precisely the length scale $z_m$ defined in (\ref{zm}). This is then the size of
the `gluonic cloud' which surrounds the quark, or in other words, the analog of the Compton wavelength for our non-Abelian source.

We will take as our starting point the results obtained in a remarkable paper
by Mikhailov \cite{mikhailov}, which we now very briefly review (more details can be found in \cite{dragtime}). This author considered an infinitely massive quark, and was able to find a solution to the
equation of motion for the dual string on AdS$_5$, for an \emph{arbitrary} timelike trajectory $x^{\mu}(\tau)$ of
the string endpoint:
\begin{equation}\label{mikhsol}
X^{\mu}(\tau,z)=z{dx^{\mu}(\tau)\over d\tau}+x^{\mu}(\tau)~.
\end{equation}
This solution is `retarded', in the sense that the behavior at
time $t=X^{0}(\tau,z)$ of the string segment located at radial position $z$ is completely
determined by the behavior of the string endpoint at an \emph{earlier} time $\tret(t,z)$ obtained
by projecting back toward the boundary along the null line at fixed $\tau$.

Using (\ref{mikhsol}),
Mikhailov was able to rewrite the total string energy in the form
\begin{equation}\label{emikh}
E(t)={\sqrt{\lambda}\over 2\pi}\int^t_{-\infty}d\tret
\frac{\vec{a}^{\,2}-\left[\vec{v}\times\vec{a}\right]^2}{\left(1-\vec{v}^{\,2}\right)^3}
+E_q(\vec{v}(t))~,
\end{equation}
with $\vec{v}\equiv d\vec{x}/dx^0$ and $\vec{a}\equiv d\vec{v}/dx^0$ the velocity and acceleration of the endpoint/quark. The first term
codifies the accumulated energy \emph{lost} by the quark
over all times prior to $t$, and, surprisingly, has the same form
as the
standard Lienard formula from classical electrodynamics. The second term in (\ref{emikh}) arises
from a total derivative on the string worldsheet, and gives the expected expression for the energy intrinsic to the quark \cite{dragtime},
\begin{equation}\label{edr}
E_q(\vec{v})={\sqrt{\lambda}\over
2\pi}\left.\left({1\over\sqrt{1-\vec{v}^{\,2}}}{1\over
z}\right)\right|^{z_m=0}_{\infty}=\gamma m~.
\end{equation}
The total spatial momentum of the string $\vec{P}$ can be similarly split \cite{mikhailov,dragtime}.
We see then that, in spite of the non-linear nature of the system, Mikhailov's procedure leads to a clean separation between the quark (including its bound field) and its gluonic radiation field.

We will now exploit this to study in more detail the dynamics of the quark with finite mass, $z_m>0$, where our non-Abelian source is no longer pointlike but has size $z_m$.
As in \cite{dragtime}, the embeddings of interest to us can be regarded as the $z\ge z_m$ portions of the Mikhailov solutions (\ref{mikhsol}). These are parametrized by (now merely auxiliary) data at the AdS boundary $z=0$, which  we will henceforth denote with tildes to distinguish them from the actual physical data associated with the endpoint/quark at $z=z_m$. In this notation, (\ref{mikhsol}) reads
\begin{equation}\label{mikhsoltilde}
X^{\mu}(\ttau,z)=z{d\tx^{\mu}(\ttau)\over d\ttau}+\tx^{\mu}(\ttau)~.
\end{equation}
Repeated differentiation with respect to $\ttau$ and evaluation at $z=z_m$ (where we can read off the quark trajectory $x^{\mu}(\ttau)\equiv X^{\mu}(\ttau,z_m)$) leads to the recursive relations
\begin{equation}\label{recursion}
{d^n x^{\mu}\over d\ttau^n}=z_m{d^{n+1}\tx^{\mu}\over d\ttau^{n+1}}+{d^n\tx^{\mu}\over d\ttau^n}\quad\forall\;n\ge 1~.
\end{equation}
Adding the $n\ge 2$ equations respectively multiplied by $(-z_m)^{n-2}$, we can deduce that
\begin{equation}\label{recursionsolution2}
{d^2 \tx^{\mu}\over d\ttau^2}={d^2 x^{\mu}\over d\ttau^2}-z_m{d^3 x^{\mu}\over d\ttau^3}+z^2_m{d^4 x^{\mu}\over d\ttau^4}-\ldots~.
\end{equation}

  Next we wish to rewrite $d\ttau$ and $d^2 \tx^{\mu}/d\ttau^2$ in terms of quantities at the actual string boundary $z=z_m$.
The first task is easy: from (\ref{mikhsoltilde}) it follows that
$$
dX^{\mu}=dz{d \tx^{\mu}\over d\ttau}+d\ttau\left(z{d^2\tx^{\mu}\over d\ttau^2}+{d \tx^{\mu}\over d\ttau}\right)~,
$$
which when evaluated at fixed $z=z_m$ implies
\begin{equation}\label{tau}
d\tau^2\equiv -dx^{\mu}dx_{\mu}=d\ttau^2\left[1-z^2_m\left({d^2\tx\over d\ttau^2}\right)^2\right]~.
\end{equation}
To arrive at this last equation, we have made use of the fact that $\ttau$ is by definition the proper time for the auxiliary worldline at $z=0$, so $(d \tx/d\ttau)^2=-1$ and $(d \tx/d\ttau)\cdot(d^2 \tx/d\ttau^2)=0$.

For the remaining task, we note first that, upon substituting the solution (\ref{mikhsoltilde}), the worldsheet momentum current evaluated at $z=z_m$ simplifies to
\begin{equation}\label{pitilde}
{2\pi\over\sqrt{\lambda}}\tPi^{z}_{\mu}={1\over z_m}{d^2 \tx_{\mu}\over d\ttau^2}+\left({d^2\tx\over d\ttau^2}\right)^2{d\tx_{\mu}\over d\ttau}~.
\end{equation}
The tilde in the left-hand side does not indicate evaluation at $z=0$ (as all other tildes do), but the fact that this current is defined as charge (momentum) flow per unit $\ttau$. The corresponding flow per unit $\tau$ is  just
$\Pi^{z}_{\mu}=(\p\ttau/\p\tau)\tPi^{z}_{\mu}$, and it is this object which according to (\ref{stringbc}) must equal the external force $\cF_{\mu}$. Using this, (\ref{tau}) and (\ref{recursion}) in (\ref{pitilde}), one can deduce  that
\begin{equation}\label{atilde}
{d^2 \tx_{\mu}\over d\ttau^2}=\frac{1}{\sqrt{1-z^4_m \slash{\cF}^2}}\left(z_m \slash{\cF}_{\mu}-z^3_m \slash{\cF}^2{d x_{\mu}\over d\tau}\right)~,
\end{equation}
where we have used the abbreviation $\slash{\cF}_{\mu}\equiv (2\pi/\sqrt{\lambda})\cF_{\mu}$.
Since $ \cF\cdot (dx/d\tau)=0$ (no work is done on the quark in its instantaneous rest frame), this implies that
$({d^2\tx/ d\ttau^2})^2=z_m^2 \slash{\cF}^2$,
which allows (\ref{tau}) to be simplified into
\begin{equation}\label{tau2}
d\ttau=\frac{d\tau}{\sqrt{1-z^4_m \slash{\cF}^2}}~.
\end{equation}

Using (\ref{atilde}) and (\ref{tau2}), we can  rewrite (\ref{recursionsolution2}) purely in terms of quark data. The result is an equation with an infinite number of higher derivatives, which we omit here for brevity.
A more manageable form of the equation of motion can be obtained by going back to (\ref{recursionsolution2}) and adding to it its $\ttau$-derivative multiplied by $z_m$, yielding
\begin{equation}
{d^2 x^{\mu}\over d\ttau^2}={d^2\tx^{\mu}\over d\ttau^2}+z_m {d^3\tx^{\mu}\over d\ttau^3}~.
\end{equation}
Through (\ref{atilde}), (\ref{tau2}) and (\ref{zm}), this can be reexpressed as
\begin{equation}\label{eom}
{d\over d\tau}\left(\frac{m{d x^{\mu}\over d\tau}-{\sqrt{\lambda}\over 2\pi m} \cF^{\mu}}{\sqrt{1-{\lambda\over 4\pi^2 m^4}\cF^2}}\right)=\frac{\cF^{\mu}-{\sqrt{\lambda}\over 2\pi m^2} \cF^2 {d x^{\mu}\over d\tau}}{1-{\lambda\over 4\pi^2 m^4}\cF^2}~,
\end{equation}
which is our main result.

This equation correctly reduces to $m d^2 x^{\mu}/d\tau^2=\cF^{\mu}$ in the pointlike limit $m\to\infty$.  Specializing to the case of motion along one dimension, it is also possible to show that (\ref{eom}) correctly reproduces the energy (and momentum) split between quark and radiation field deduced at finite $m$ in \cite{dragtime}, and in fact
 encodes the Lorentz-covariant generalization of that split. To make this explicit, we rewrite (\ref{eom}) in the form
 \begin{equation}\label{eomsplit}
 {d P^{\mu}\over d\tau}\equiv {d p_q^{\mu}\over d\tau}+{d P^{\mu}_{\mbox{\scriptsize rad}}\over d\tau}=\cF^{\mu},
 \end{equation}
 recognizing $P^{\mu}$ as the total string ($=$ quark $+$ radiation) four-momentum,
 \begin{equation}\label{pq}
 p_q^{\mu}=\frac{m{d x^{\mu}\over d\tau}-{\sqrt{\lambda}\over 2\pi m} \cF^{\mu}}{\sqrt{1-{\lambda\over 4\pi^2 m^4}\cF^2}}
 \end{equation}
 as the intrisic momentum of the quark, and
\begin{equation}\label{radiationrate}
{d P^{\mu}_{\mbox{\scriptsize rad}}\over d\tau}={\sqrt{\lambda}\, \cF^2 \over 2\pi m^2}\left(\frac{{d x^{\mu}\over d\tau}-{\sqrt{\lambda}\over 2\pi m^2} \cF^{\mu} }{1-{\lambda\over 4\pi^2 m^4}\cF^2}\right)
\end{equation}
as the rate at which momentum is carried away from the quark by chromo-electromagnetic radiation. (In some cases,
this vacuum radiation rate can also be relevant for quark motion within a thermal plasma \cite{dragtime,vacinplasma}.)

We can immediately deduce from (\ref{pq}) the mass-shell condition $p_q^2=-m^2$, which shows in particular that the split $P^{\mu}=p_q^{\mu}+P^{\mu}_{\mbox{\scriptsize rad}}$ defined in (\ref{eomsplit})-(\ref{radiationrate}) is correctly Lorentz covariant. As we indicated above, $P^{\mu}_{\mbox{\scriptsize rad}}$
  represents the portion of the total four-momentum stored at any given time in the \emph{purely radiative} part of the gluonic field set up by the quark. The remainder, $p^{\mu}_q$, includes the contribution of the \emph{bound} field sourced by our particle, or in quantum mechanical language, of the gluonic cloud surrounding the quark, which gives rise to the deformed dispersion relation seen in (\ref{pq}). In other words, $p^{\mu}_q$ is the four-momentum of the composite quark. Surprisingly, all of this is closely analogous to the classical electromagnetic case, and in particular, to the covariant splitting of the energy-momentum tensor achieved in \cite{teitelboim}.

As noticed already in \cite{dragtime} for the case of linear motion,
  a prominent feature of the equation of motion (\ref{eom}), as well as the dispersion relation (\ref{pq}) and radiation rate (\ref{radiationrate}), is the presence of a divergence when $\cF^2=\cF^2_{\mbox{\scriptsize crit}}$, where
 $
  \cF^2_{\mbox{\scriptsize crit}}={ 4\pi^2 m^4/\lambda}
 $
  is the critical value at which the force becomes strong enough to nucleate quark-antiquark pairs (or, in dual language, to create open strings). 

Let us now examine the behavior of a quark that is sufficiently heavy, or is forced sufficiently softly, that the condition $\sqrt{\lambda |\cF^2|}/2\pi m^2\ll 1$ (i.e., $|\cF^2|\ll |\cF^2_{\mbox{\scriptsize crit}}|$) holds. It is then natural to expand the equation of motion in a power series in this small parameter. To zeroth order in this expansion, we have the pointlike result
$m \p_{\tau}^2 x^{\mu}=\cF^{\mu}$,
as we had already mentioned above. If we instead keep terms up to first order, we find
  $$
  m {d\over d\tau}\left( {d x^{\mu}\over d\tau}-{\sqrt{\lambda}\over 2\pi m^2}\cF^{\mu}\right)=\cF^{\mu}-{\sqrt{\lambda}\over 2\pi m^2}\cF^2 {d x^{\mu}\over d\tau}~.
  $$
  In the $\cO(\sqrt{\lambda})$ terms it is consistent, to this order, to replace $\cF^{\mu}$ with its zeroth order value, thereby obtaining
  \begin{equation}\label{ourald}
  m \left( {d^2 x^{\mu}\over d\tau^2}-{\sqrt{\lambda}\over 2\pi m}{d^3 x^{\mu}\over d\tau^3}\right)=\cF^{\mu}-{\sqrt{\lambda}\over 2\pi}{d^2 x^{\nu}\over d\tau^2}{d^2 x_{\nu}\over d\tau^2} {d x^{\mu}\over d\tau}~.
  \end{equation}
  Interestingly, this coincides \emph{exactly} with the Lorentz-Dirac equation (\ref{ald}).  As expected from the preceding discussion, on the left-hand side we find the Schott term (associated with the bound field of the quark) arising from the modified dispersion relation (\ref{pq}). On the right-hand side we see the radiation reaction force given by the covariant Lienard formula, as expected from the  result (\ref{emikh}) \cite{mikhailov}, which is the pointlike limit of the radiation rate (\ref{radiationrate}).
  Moreover, by comparing (\ref{ald}) and (\ref{ourald}) we learn that it is the Compton wavelength (\ref{zm}) that plays the role of characteristic size $r_e$ for the composite quark. This is indeed the natural quantum scale of the problem.

 We can continue this expansion procedure to arbitrarily high order in $\sqrt{\lambda |\cF^2|}/2\pi m^2$. At order $n$ in this parameter, we would obtain an equation with derivatives up to order $n+2$.
Our full equation (\ref{eom}) is thus recognized as an extension of the Lorentz-Dirac equation that automatically incorporates the size $z_m$ of our non-classical,
non-pointlike and non-Abelian source.

It is curious to note that (\ref{eom}), which incorporates the effect of radiation damping on the quark, has been obtained from (\ref{mikhsol}), which does \emph{not} include such damping for the string itself. The supergravity fields set up by the string are of order $1/N_c^2$, and therefore subleading at large $N_c$. Even more curious
\cite{cg} is the fact that it is precisely these suppressed fields that encode the gluonic field profile generated by the quark, as has been explored in great detail in recent years  \cite{gubserreview}. It would be interesting to explore how the split into bound and radiation fields is achieved from this perspective.

The passage from (\ref{ourald}) to (\ref{eom}), which can be viewed intuitively as the addition of an infinite number of higher derivative terms $(z_m d/d\tau)^n$, has a profound impact on the space of solutions. Here we will limit ourselves to two general observations, leaving the search for specific examples of solutions to a more extensive report \cite{damping}. The first is to notice that, unlike its classical electrodynamic counterpart  (\ref{ald}), our composite quark equation of motion has no pre-accelerating or self-accelerating solutions. In the (continuous) absence of an external force, (\ref{eom}) uniquely predicts that the four-acceleration of the quark must vanish. Our second observation, however, is that the converse to this last statement is not true: constant four-velocity does not uniquely imply a vanishing force. This is again a consequence of the extended, and hence deformable, nature of the quark.

All in all, then, we have in (\ref{eom}) a physically sensible and interesting description of the dynamics of a composite quark in SYM. It is truly remarkable that the AdS/CFT correspondence grants us such direct access to this piece of strongly-coupled non-Abelian physics.

\noindent
{\bf Acknowledgments.} We thank David Mateos and Mat\'\i as Moreno for useful discussions. This work was  supported by Mexico's National Council of Science and Technology (CONACyT) grant 50-155I and DGAPA-UNAM grant IN116408.


\begin{thebibliography}{99}
\bibitem{foot}
The relevant background is actually 
AdS$_5\times\bS^5$, with a constant dilaton and $N_c$ units of Ramond-Ramond five-form flux through the five-sphere, but these additional features will play no direct role in our analysis.
\end{thebibliography}
\end{document}